\documentstyle[12pt,epsf,epsfig]{article}
\def\ga{\mathrel{\raise.3ex\hbox{$>$\kern-.75em\lower1ex\hbox{$\sim$}}}}
\def\la{\mathrel{\raise.3ex\hbox{$<$\kern-.75em\lower1ex\hbox{$\sim$}}}}
\def\gyr{{\rm \, G\kern-0.125em yr}}
\def\gev{{\rm \, Ge\kern-0.125em V}}
\def\tev{{\rm \, Te\kern-0.125em V}}
\def\beq{\begin{equation}}
\def\eeq{\end{equation}}
\def\ss{\scriptscriptstyle}
\def\scs{\scriptstyle}
\def\mb{m_{\widetilde B}}

\def\mchi{m_{\tilde \chi}}
\def\msf{m_{\tilde f}}
\def\m12{m_{1\!/2}}

\def\mf{m_{\ss{f}}}

\def\ohsq{\Omega_{\widetilde\chi}\, h^2}
\def\ch{\widetilde \chi}
\def\st{{\widetilde \tau}_{\scriptscriptstyle\rm R}}
\def\sm{{\widetilde \mu}_{\scriptscriptstyle\rm R}}
\def\sel{{\widetilde e}_{\scriptscriptstyle\rm R}}
\def\sl{{\widetilde \ell}_{\scriptscriptstyle\rm R}}

\begin{document}
\begin{titlepage}
\pagestyle{empty}
\baselineskip=21pt
\rightline{hep-ph/9810360}
\rightline{CERN-TH/98-326}
\rightline{UMN--TH--1725/98}
\rightline{MADPH-98-1087}
\rightline{October 1998}
\vskip.25in
\begin{center}

{\large{\bf Neutralino-Stau Coannihilation and 
 the Cosmological Upper Limit on the Mass of the Lightest Supersymmetric 
Particle}}
\end{center}
\begin{center}
\vskip 0.5in
{John Ellis}

{\it Theory Division, CERN, CH-1211 Geneva 23, Switzerland}

{Toby Falk}

{\it Department of Physics, University of Wisconsin, Madison, WI~53706,
USA}

 and 

Keith A.~Olive

{\it
{School of Physics and Astronomy,
University of Minnesota, Minneapolis, MN 55455, USA}\\}
\vskip 0.5in
{\bf Abstract}
\end{center}
\baselineskip=18pt \noindent
We consider the effects of neutralino-stau ($\chi - {\tilde \tau}$) 
coannihilations on the
cosmological relic density of the lightest supersymmetric particle (LSP)
$\ch$ in the minimal supersymmetric extension of the Standard Model
(MSSM), particularly in the constrained MSSM in which
universal supergravity inputs at the GUT scale are assumed.
For much of the parameter space in these models, $\ch$ is
approximately a $U(1)$ gaugino ${\tilde B}$,
and constraints on the cosmological relic density $\Omega_{\tilde B} h^2$
yield an upper bound on $\mb$.
We show that in regions of
parameter space for which the cosmological bound is nearly saturated,
coannihilations of the $\tilde B$ with the ${\tilde \tau}$, the next
lightest sparticle, are important  and may reduce significantly the
$\tilde B$ relic density.  Including also ${\tilde B}$ coannihilations
with the $\tilde e$ and
$\tilde \mu$, we find that
the upper limit on $m_{\ch}$ is increased from
about $200$~GeV to about $600$~GeV in the constrained MSSM, 
with a similar new upper limit expected in the MSSM.
\end{titlepage}
\baselineskip=18pt

Supersymmetry is one of the most appealing options for possible
physics beyond the Standard Model, motivated theoretically by the
help it offers in stabilizing the gauge hierarchy, its successful
prediction of sin$^2 \theta_W$ in the context of GUTs, and its
consistency with the range of Higgs boson masses favoured by the 
precision electroweak data. Accordingly, supersymmetry has been the
focus of intense phenomenological studies, particularly in the
framework of the minimal supersymmetric extension of the Standard
Model (MSSM)~\cite{MSSM}. Many of these studies assume that supersymmetry
is
broken by unspecified dynamics in some hidden sector of the theory,
which is communicated to the observable MSSM particles by gravitational
interactions. One may further assume that the supersymmetry-breaking
mass parameters $m_0,m_{1/2}, A$ and $B$ are universal at
the supergravity input scale, providing the constrained MSSM (CMSSM)
framework that is privileged in this paper.

The lightest supersymmetric particle (LSP) is stable in the MSSM, unless
additional $R$-violating interactions are postulated~\cite{pf}. The LSP
is 
generally thought to be the lightest neutralino $\ch$~\cite{hg,ehnos} and
is a
favoured candidate for the Cold Dark Matter favoured by astrophysicists
and theorists of structure formation. The relic LSP density can in
principle be calculated reliably as a function of the parameters
of the MSSM~\cite{swo}. It is remarkable that there is
a large generic domain of the parameter space in the MSSM and in the
CMSSM, consistent with all experimental constraints~\cite{efgos},
in which the $\ch$ has a relic mass
density $\Omega_{\ch} h^2 \sim 0.1$ as favoured by astrophysical
and cosmological arguments.

The phenomenological arguments for supersymmetry based on the
gauge hierarchy, sin$^2 \theta_W$ and the Higgs mass $m_h$ all suggest that
supersymmetric particles should weigh $\sim 1$~TeV or less, but do
not provide very precise upper limits on their masses. For example,
the amount of fine tuning required to maintain the gauge hierarchy
increases as the MSSM mass parameters are increased, but there is no
objective criterion how much fine tuning is tolerable~\cite{ceop}.
Moreover,
sin$^2 \theta_W$ and $m_h$ are only logarithmically sensitive to the
sparticle masses. On the other hand, the LSP relic mass density is
very sensitive to $m_{\ch}$, since the annihilation cross section
tends to decrease as $m_{\ch}$ increases, increasing
also its relic number number density. The relic density is also very
sensitive over much of the parameter space to the scalar mass parameters,
as these, too, control the annihilation cross section. The constraints on
the general MSSM parameter space have been explored in some detail,
particularly in the CMSSM framework. The possibility that the LSP might
be mainly a photino
${\tilde \gamma}$ has been excluded by lower limits on sparticle masses
from LEP and elsewhere~\cite{efgos}. When studying the possibility that
the LSP might be largely a Higgsino ${\tilde H}$,
coannihilations~\cite{gs} between the Higgsino-like LSP and the
next-to-lightest supersymmetric particle (NLSP) have to be taken into
account~\cite{co2,dn}. This Higgsino LSP possibility has also been
tightly constrained by LEP and may be explored completely by upcoming
runs~\cite{efgos}.
This leaves us with the likelihood of a Bino- (${\tilde B}$-) like
LSP, and cosmology imposes an important upper limit on its mass, which
has been given as $m_{\tilde B} \la 300$~GeV in the 
MSSM \cite{up} and $ \la 200$~GeV the CMSSM framework \cite{upper}.

The purpose of this paper is re-evaluate this upper limit,
including for the first time detailed calculations of coannihilations
between the ${\tilde B}$ LSP and the lighter supersymmetric partner of the
$\tau$, the right-handed stau ${\tilde \tau}_R$, which is the NLSP
in much of the ${\tilde B}$ LSP region. We find that 
${\tilde B} - {\tilde \tau}_R$ coannihilation is particularly important
when the relic mass density is close to the cosmological upper limit,
which we take to be $\Omega_{\ch} h^2 < 0.3$, resulting in a
considerable relaxation of the previous upper bound on $m_{\tilde B}$.
Including also $\tilde B$ coannihilation with the ${\tilde e}_R$ and
${\tilde \mu}_R$, we now
find that $m_{\tilde B}$ may be as large as $600$~GeV in the CMSSM
or MSSM.

As already commented, the LSP is a ${\tilde B}$
in much of parameter space that leads to an interesting relic density,
both in the generic MSSM and in the CMSSM~\cite{efgos}.
Indeed, this is a prediction of the CMSSM.
Unless the ${\tilde B}$ mass happens to lie near $m_Z/2$
or $m_h/2$, in which case there are large contributions to the
annihilation through direct $s$-channel resonance exchange, the dominant
contribution to
the $\tilde{B} \tilde{B}$ annihilation cross section comes from crossed
$t$-channel sfermion exchange. The resonant case is anyway not relevant
for the upper bounds on $m_{\tilde B}$ to be discussed here.
In the absence of such a resonance, the thermally-averaged cross section
for $\tilde{B} \tilde{B} \to f \bar{f}$ takes the generic form
\begin{equation}
\langle \sigma v \rangle = (1 - {\mf^2 \over \mb^2})^{1/2} {g_1^4 \over
128 \pi} \left[ (Y_L^2 + Y_R^2)^2 ({\mf^2 \over \Delta_f^2}) \,+\, (Y_L^4 +
Y_R^4)  ({4 \mb^2 \over \Delta_f^2}) (1 + ...) \,x \,\right]
\label{eqn:sigv}
\end{equation}
where $Y_{L(R)}$ is the hypercharge of $f_{L(R)}$, $\Delta_f \equiv
\msf^2 + \mb^2 - \mf^2$, and we have shown only the leading $P$-wave
contribution proportional to $x \equiv T/\mb$.

The upper limit on $m_{\tilde B}$ due to the cosmological
relic density comes about as follows~\cite{up}.  The assumption that
the ${\tilde B}$ is
the LSP requires, in particular, that $\mb < \msf$. 
In order to minimize the relic
density, we must maximize the cross section, which is done by setting
$\msf = \mb$.  The cross section is then approximately inversely
proportional to $\mb^2$. The 
cosmological upper limit on $\Omega_{\tilde B}
h^2$ translates into a lower limit on $\langle \sigma v \rangle$ which 
then, in turn, yields an upper limit to $\mb$.  In the MSSM, this limit is
$\mb \la
300$ GeV, when all sfermion masses are taken to be equal at the weak
scale.

In the CMSSM, the argument is somewhat similar, although $\mb$
and the sfermion masses are no longer entirely
independent, because it is assumed in the CMSSM
that there is a common scalar
mass $m_0$ at the GUT scale. For a given value of the common
gaugino
mass $\m12$ at the GUT scale, the relic ${\tilde B}$ density falls with
$m_0$, since 
$\msf^2=m_0^{2} + C_{f} \m12^2 + O(m_Z^2)$, where $C_f$ is a 
positive numerical
coefficient that is calculable via the renormalization-group evolution of
the sfermion masses.
Therefore, the cosmological upper limit on $\Omega_{\tilde B} h^2$
translates at fixed $\m12$ into an upper limit on $m_0$.
Typically, this upper limit is not larger than $m_0 \la 150$ GeV, unless
one is sitting on a direct-channel pole, i.e., when $\mb \sim m_Z/2$ or
$m_h/2$, in which case
$s$-channel annihilation is dominant and there is no upper limit
to $m_0$. However, as already mentioned, this is not our case, as we are
interested in an upper bound on $\mb$. We recall that $\mb$ scales with
$\m12$, and it transpires for $\m12 \ga 400$ GeV that
$\mb$  exceeds mass of the lightest sfermion, which is
typically the ${\tilde \tau}_R$, for $m_0$ small enough to satisfy the
cosmological bound \cite{upper}. Thus, the LSP is no longer a neutralino
for such large values of $\m12$, and
hence an upper bound $\mb \la 200$~GeV~\cite{upper} can be
established.~\footnote{This upper
bound can be strengthened by requiring that the global minimum
of the effective potential of the MSSM conserve electric charge and
color~\cite{af}.}

When $\mb$ attains this upper bound, the ${\tilde B}$
is degenerate in mass with the
${\tilde \tau}_R$, and quite close in mass to the ${\tilde e}_R$ and
${\tilde
\mu}_R$. It is well known~\cite{gs} that, in such circumstances,
the neutralinos can be maintained in equilibrium by scatterings with a
slightly heavier particle, and the number density of neutralinos can be
significantly reduced by such coannihilations.  The case
of heavy Higgsinos is a well studied example~\cite{dn}. 
Analogously to that case,
the ${\tilde B}$ relic density can be reduced through coannihilation
with slightly heavier ${\st}$'s or other sleptons,
as we now discuss in detail.

To derive a thermally-averaged
cross section, we use the technique of~\cite{swo}. Thus, we expand
$\langle\sigma v_{\rm rel}\rangle$ in a Taylor expansion in powers of
$x = T/\mb$:
\begin{equation}
\label{d}
\langle \sigma v_{\rm rel } \rangle = a + b x + O(x^2)\;.
\end{equation}
Repeating the analysis~\cite{swo} for initial particles with 
different masses $m_1$ and $m_2$ yields
\begin{equation}
\langle\sigma v_{\rm rel}\rangle = \frac{1}{m_1m_2}\left(1 -
 \frac{3(m_1 + m_2)T}{2 m_1m_2}\right) w(s)|_{s\to (m_1+m_2)^2 + 3(m_1+m_2)T}
\; +\; O(T^2).
\label{e}
\end{equation}
where
\begin{eqnarray}
w(s) & \equiv & \frac{1}{4}\int\; d{\rm LIPS}\; |{\cal M}|^2\\
\noalign{\medskip}
     & = & \frac{1}{32\pi}\frac{p(s)}{s^{1/2}}\int^{+1}_{-1}
            d\cos\theta_{\rm CM}\;|{\cal M}|^2,
\end{eqnarray}
Here $d{\rm LIPS}$ is the Lorentz-Invariant phase-space element,
$p(s)$ is the magnitude of the three momentum of one of the initial
particles in the center-of-mass frame, as a
function of the total center-of-mass energy-squared $s$,  
$\theta_{\rm CM}$ is the center-of-mass scattering angle, and 
$|{\cal M}|^2$ is the absolute square of the reduced matrix element for
the annihilation, summed over final spins and averaged over initial spins
wherever appropriate.  The $a$ and $b$ coefficients in (\ref{d})
may be read off the right-hand side of (\ref{e}), after expanding
in powers of $x$.~\footnote{A similar calculation is necessary when the
LSP is assumed to be a sneutrino \cite{fkosi}.}

If the masses of the next-to-lightest sparticles (NLSPs) are close to
the LSP mass: $\Delta M = {\cal O}(x_f) M$, where $x_f \sim (1/20-1/25)$
is
the value of $T/\mchi$ at the time of neutralino decoupling, 
the number densities of the NLSPs have only slight
Boltzmann suppressions with respect to the LSP number density when the
LSP freezes out of chemical equilibrium with the thermal bath. In such
a case, coannihilations of NLSPs with the LSP, along with NLSP-NLSP
annihilations, may play an important r\^ole in keeping the LSPs in
chemical equilibrium with the bath~\cite{gs}.  These processes can be
particularly important when the LSP annihilation rate itself is
suppressed, as is the case for neutralinos. Gaugino-like
neutralinos typically annihilate predominantly into fermion pairs, 
and such processes
exhibit $P$-wave suppressions~\cite{hg}, so that $a\ll b$ in (\ref{d}).
This effect can be seen from (\ref{eqn:sigv}) where the $a$-term
is suppressed relative to $b$ by $m_f^2/\mb^2$,
reducing the neutralino annihilation cross section by a factor
of ${\cal O}(x_f)$. We also emphasize that 2-2
scatterings with particles in the thermal bath keep the NLSPs, in this
case the $\st$, $\sel$ and $\sm$, in chemical equilibrium with each
other and with $\ch$, down to temperatures well below the temperature at which 
the comoving LSP number density freezes out.

We consider the total density $n \equiv \sum_i n_i$, where the index
$i$ runs
over $\st, \st^*, \sel, \sel^*, \sm$ and  $\sm^*$ as well as $\ch$, and
write the rate equation for $n$:
\begin{equation}
\frac{dn}{dt}  =  -3Hn - \langle \sigma_{\rm eff} v_{\rm rel } \rangle 
                    (n^2 - n_{\rm eq}^2),
\label{b}
\end{equation}
where $H$ is the Hubble parameter, and
\begin{equation}
\sigma_{\rm eff} = \sum_{ij}\sigma_{ij} r_i r_j.
\end{equation}
Here $r_i \equiv n_i^{\rm eq}/n^{\rm eq}$ where
$n^{\rm eq}_i$ is the equilibrium density of particle species $i$, and
$\sigma_{ij}$ is the total cross section for particle $i$
to annihilate with particle $j$. Since
the sleptons decay into neutralinos after freeze-out, the number
density of neutralinos becomes $n$.   Many  of the $\sigma_{ij}$ are
related, and if we take the
$\sel$ and $\sm$ to be degenerate in mass and ignore the electron and
muon mass, we can write
\begin{eqnarray}
\sigma_{\rm eff} &=& \sigma_{\chi\chi} r_{\chi}r_{\chi} + 4\, \sigma_{\chi\tau} r_{\chi} r_{\tau} +
8\,  \sigma_{\chi e} r_{\chi} r_{e} +2\, (\sigma_{\tau\tau}+\sigma_{\tau\tau^*})r_{\tau}r_{\tau}+8\, (\sigma_{\tau e}+\sigma_{\tau e^*})r_{\tau}r_{e} + \hfill\nonumber \\
&&4\,  (\sigma_{e e}+\sigma_{e e^*})r_{e}r_{e}+ 4\,  (\sigma_{e \mu}+\sigma_{e \mu^*})r_{e}r_{e}
\end{eqnarray}
In Table~\ref{table:states} we list the sets of initial and final
states for which we compute  the annihilation cross sections.  The
cross sections for other reactions, such as
$\sl\sl^*\rightarrow hH, hA, hZ, H^+H^-,$ \\
$W^+H^-, \ldots$ are either
suppressed or kinematically unavailable in the regions of CMSSM
parameter space relevant to our analysis.  In practice, we find that the dominant
contributions to $\sigma_{\rm eff}$ come from annihilations of
$\sl^{\,i}\,\sl^{\,i^{\scs *}}$ to gauge bosons,
$\sl^{\,i}\,\sl^{\,j}$ to lepton pairs, and $\sl^{\,i}\,\ch$ to
$\ell^i \;+\,$ gauge boson.

\begin{table}[htb]\caption{Initial and Final States for Coannihilation:
$\{i,j=\tau,e,\mu\}$}
\begin{center}
\begin{tabular}{c|l}\hline
Initial State & Final States\\ \hline\\
$\sl^{\,i}\,\sl^{\,i^{\scs *}}$    & $\gamma\gamma,\, \, ZZ\, ,\,\gamma Z,\, W^+W^-, \,h h, \,\ell^{\,i}\bar \ell^{\,i}$\\[0.5ex]
$\sl^{\,i}\,\sl^{\,j}$  & $\ell^{\,i}\ell^{\,j}$\\[0.5ex]
$\sl^{\,i}\,\sl^{\,{j}^{\scs *}},\,i\neq j$   & $\ell^{\,i}\bar \ell^{\,j}$\\[0.5ex]
$\sl^{\,i}\,\ch$   & $\ell^{\,i}\gamma, \ell^{\,i}Z, \ell^{\,i}h$\\[0.5ex]
\label{table:states}
\end{tabular}
\end{center}
\end{table}

To get a simple estimate for the size of the effect of including the
next-to-lightest states, we first assume degenerate LSP and NLSPs, and
consider a model in which the NLSP-NLSP and NLSP-LSP annihilations are
all unsuppressed.  Thus, we take 
\begin{equation}
\label{ab}
\{a_{ij}\approx a_{1j}\approx b_{11},a_{11}=0; i,j > 1\}, 
\end{equation}
where the subscripts $i,j$ refer to the NLSPs and 1 to the LSP.
Denoting with superscripts $0$ quantities that are
computed ignoring the NLSP states, we estimate the following
ratio of relic densities without and with coannihilation:
\begin{equation}
R\equiv{\Omega^0\over\Omega} \approx ({2\over x_{\!\ss f}^{\scs0}})\,({a_{\rm eff}\over b_{11}})\,
({x_{\!\ss f}\over x_{\!\ss f}^{0}}),
\end{equation}
where ${x_{\!\ss f}^0 / x_{\!\ss f}}\approx 1+x_{\!\ss f}^0 \ln
(g_{\rm tot}/g_1 x_{\!\ss f}^0)\approx 1.2$, $g_{\rm tot}=\sum_i g_i$,
and ${a_{\rm eff}/ b_{11}}\approx 1-g_1^2/g_{\rm tot}^2=15/16$ for the
case of three degenerate slepton NLSPs.  Thus, in this crude approximation
we find a factor $\sim 35$ reduction in the relic density.   Ignoring
the 
(heavier) left-handed sleptons, we may reduce (\ref{eqn:sigv})
to $\langle \sigma v \rangle \approx 3 g_1^4 x/(16\pi \mchi^2) $, 
yielding $\ohsq\sim8\times10^{-6}\;\mchi^2$.  Thus,
in this simple approximation, $\ohsq=(\ohsq)^0/R<0.3$
gives an  upper bound on the ${\tilde B}$ mass of $\mb \la1.2\tev$.

\begin{figure}
\vspace*{-1.8in}
\begin{minipage}{6.0cm}
\hspace*{-1in}
\epsfig{file=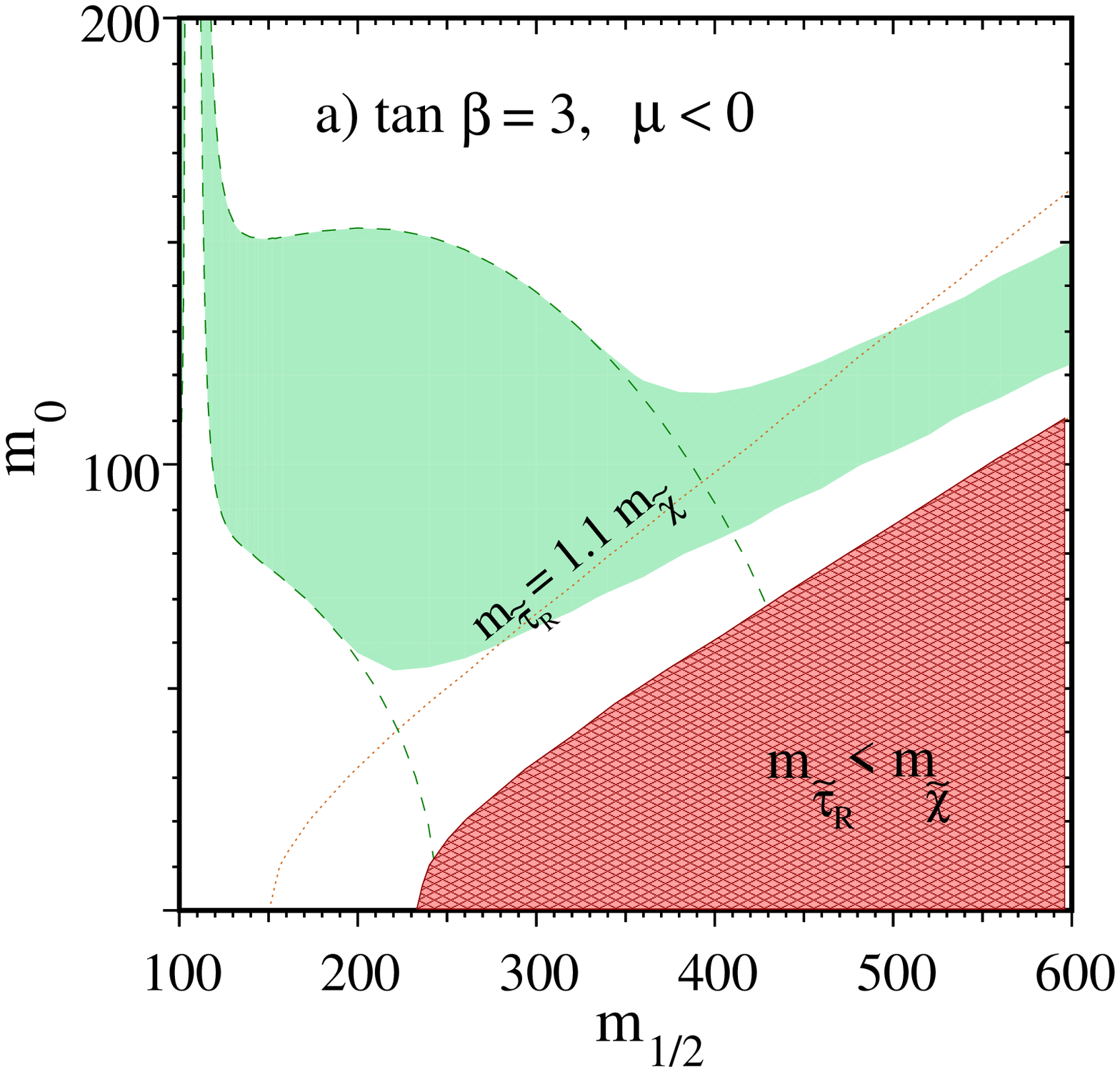,height=6in} 
\end{minipage}
\hspace*{0.3in}
\begin{minipage}{6.0cm}
\epsfig{file=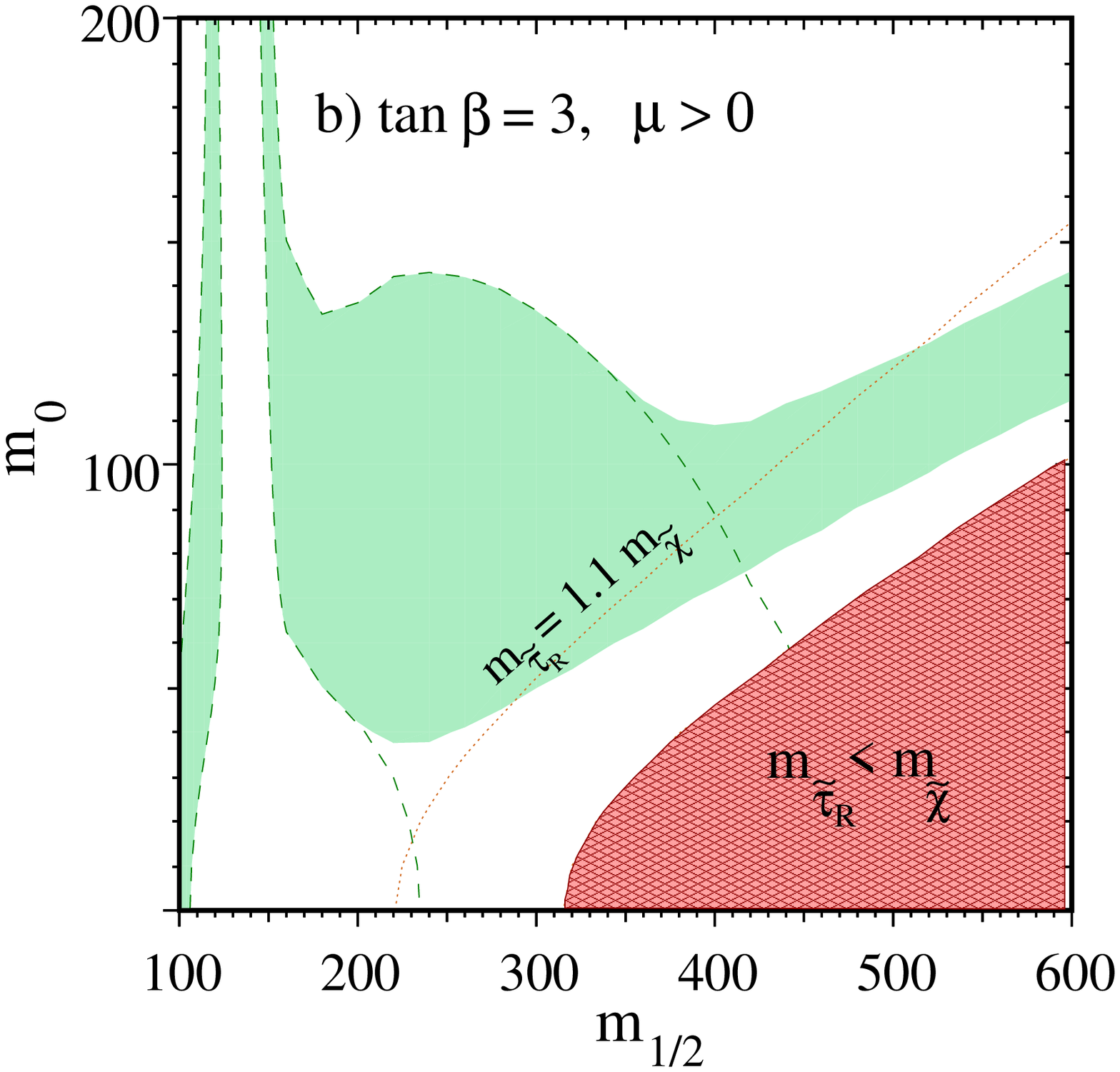,height=6in} 
\end{minipage}\hfill
\vspace{-2.0in}
\begin{minipage}{6.0cm}
\hspace*{-1in}
\epsfig{file=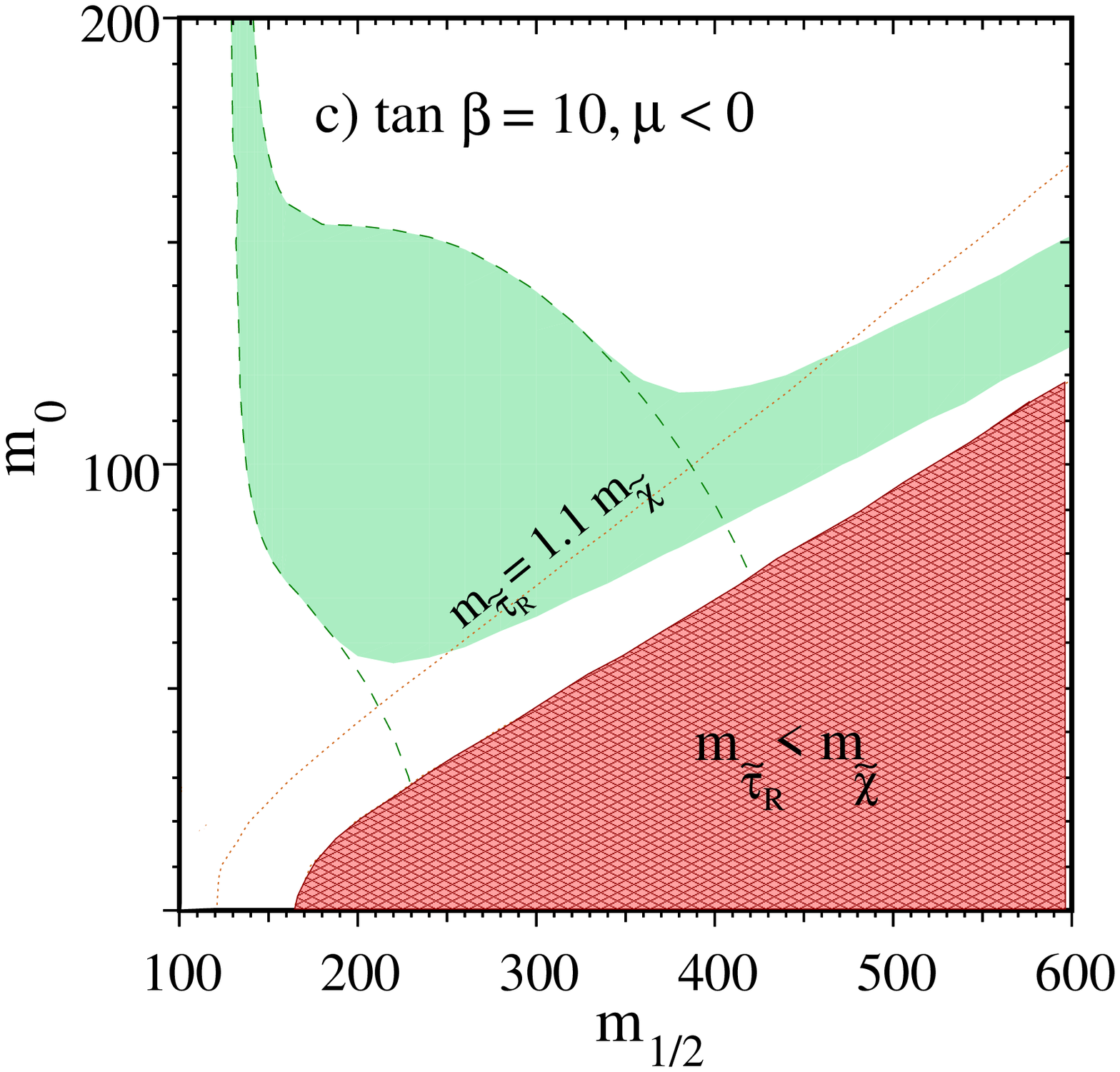,height=6in} 
\end{minipage}
\hspace*{0.3in}
\begin{minipage}{6.0cm}
\epsfig{file=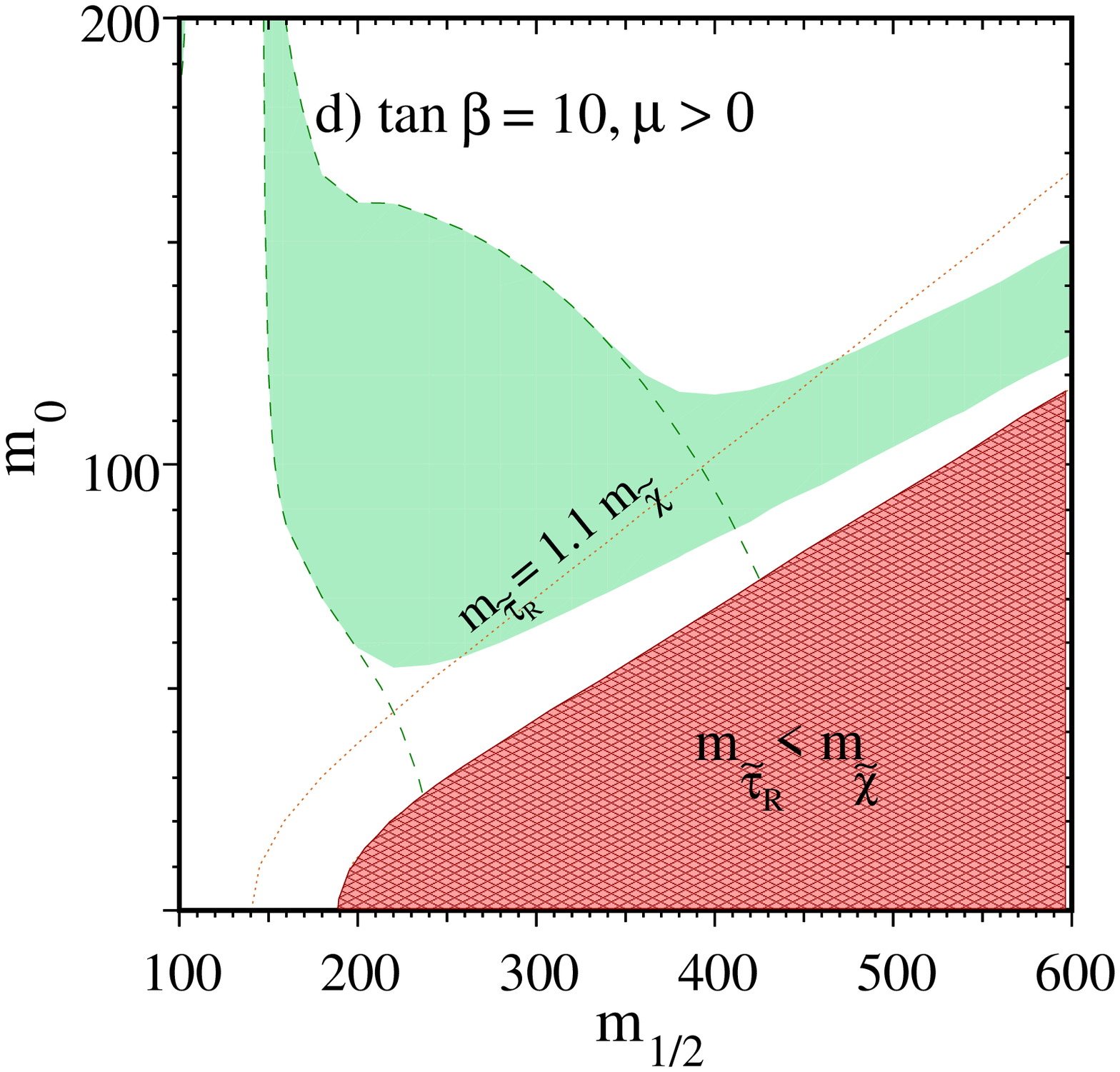,height=6in} 
\end{minipage}\hfill
\vskip-0.7in
\caption{\label{fig:sm}The light-shaded area is the cosmologically preferred 
region with \mbox{$0.1\leq\ohsq\leq 0.3$}.   The dashed line shows the location of the
cosmologically preferred region  if one ignores the light sleptons.  
In the dark shaded
region in the bottom right of each panel, the LSP is the ${\tilde
\tau}_R$, leading to an unacceptable abundance
of charged dark matter.  Also shown as a dotted line is the
contour $m_{\st}=1.1 \,\mchi$.}
\end{figure}

\begin{figure}
\vspace*{-2.3in}
\begin{minipage}{6.0cm}
\hspace*{-1in}
\epsfig{file=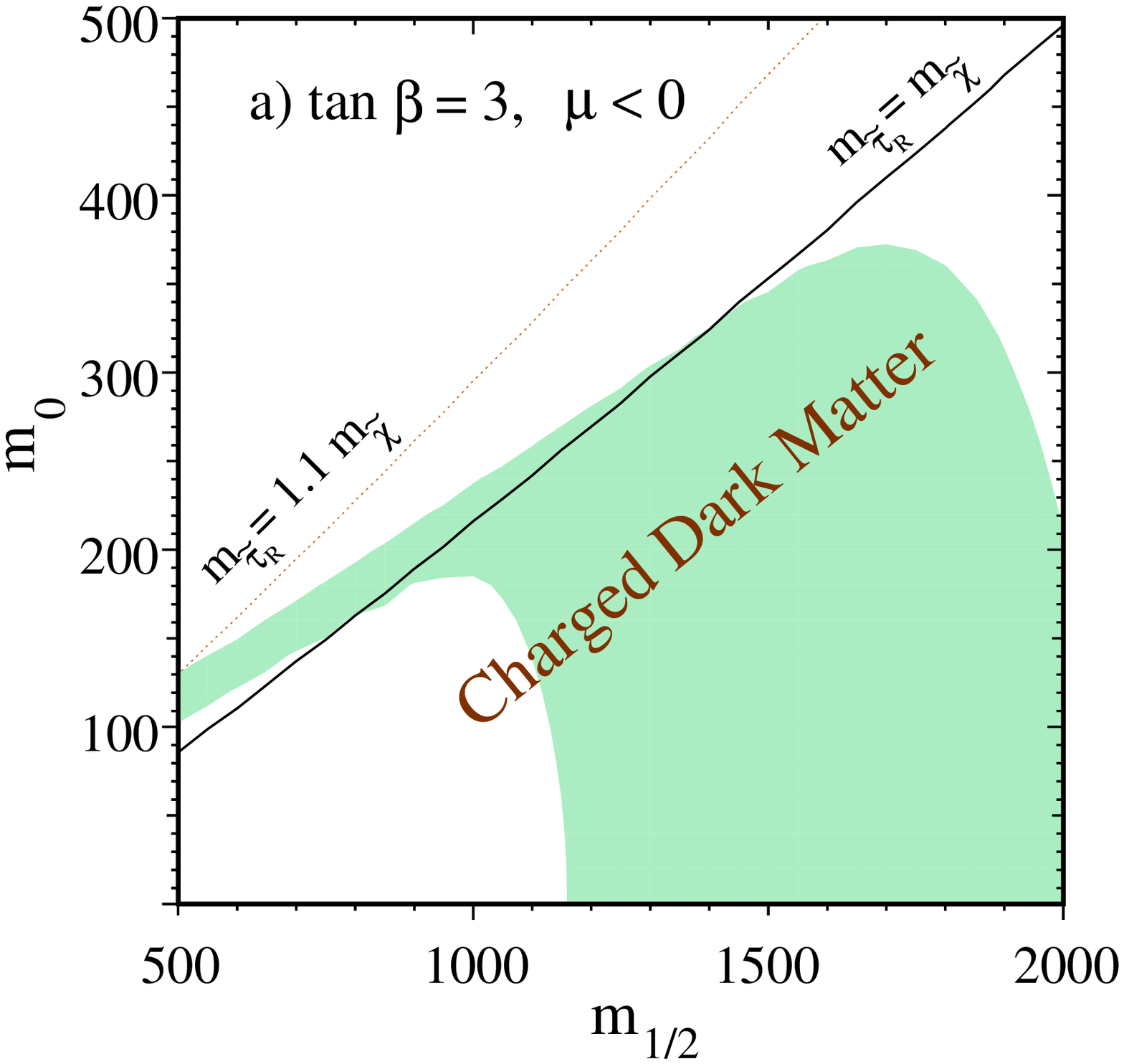,height=6in} 
\end{minipage}
\hspace*{0.3in}
\begin{minipage}{6.0cm}
\epsfig{file=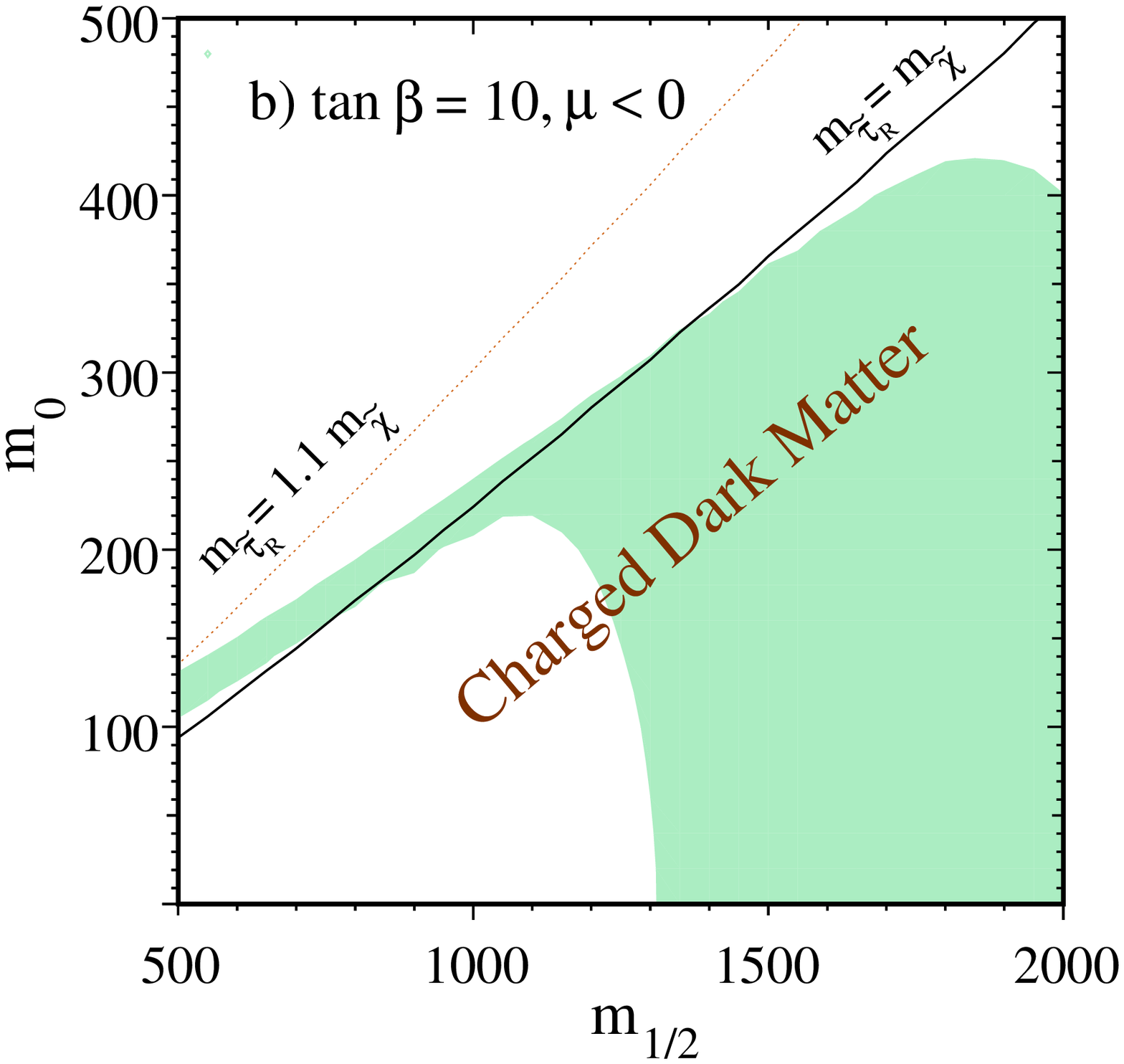,height=6in} 
\end{minipage}\hfill
\caption{\label{fig:bg}Same as Fig.~\protect{\ref{fig:sm}}(a,c),  
extended to larger values of $\m12$.}
\end{figure}

We have gone beyond the above crude approximations to make a detailed
numerical analysis of coannihilation
effects on the neutralino relic density, including light sleptons,
some of whose results are displayed in Fig.~\ref{fig:sm}.
The light
shaded region corresponds to \mbox{$0.1<\ohsq<0.3$}, and the dark shaded
region to $m_{\st}<\mchi$.  We have chosen the representative points
$\tan\beta=3$ and 10, and present results for both $\mu<0$ and $\mu>0$.
In practice,
the relationship (\ref{ab}) is not exact, not all of the $a_{ij}$ are
unsuppressed, there are contributions from the $b_{ij}$, and the
$\sel$ and $\sm$ can be slightly heavier than the $\st$. These
corrections 
have the net effect of reducing $\sigma_{\rm eff}$ by a factor of
$\sim(3-4)$.  Numerically, we find that $R\sim10$ along the line where
$\mchi=m_{\st}$, at the top of the dark shaded region in
Fig.~\ref{fig:sm}.  As $m_0$ increases, the
sleptons become heavier relative to the neutralino, and their number
density rapidly falls, reducing their contribution to $\sigma_{\rm
  eff}$.  The relic density rises rapidly in this region, leaving an
allowed band in $m_0$ which is about 30-50 $\gev$ wide for $\m12 <
800\gev$.  In Fig.~\ref{fig:bg} we extend the coverage of
Fig.~\ref{fig:sm}a,c over a
larger scale, to show the cross-over point between the regions with
$\ohsq<0.3$ and $m_{\st}<\mchi$, where there would be an unacceptable
abundance of charged dark matter~\cite{ehnos}.  The two constraints
together require $\m12\la1450$, corresponding to an upper bound on the
neutralino mass of $\mchi\la600\gev$. The
results for $\mu>0$ and $\mu < 0$ are very similar, so we do not display
the latter.  The width of the allowed region is insensitive to $A_0$, though the 
position of the line $\mchi=m_{\st}$ can vary somewhat.
As already commented, the requirement that the electroweak
vacuum conserve electric charge and color constrains significantly the
CMSSM parameter space~\cite{af}. We find that 
the large-$\m12$ tail of the
region newly allowed by coannihilation obeys this requirement, for
$\tan\beta \ga 3$ for some values of $A_0$.

We expect the corresponding bound in the MSSM to be very similar.  In
the general case, one must take all the squarks and sleptons
degenerate with the neutralino and compute the annihilation and
coannihilation cross-sections for all possible combination of
sfermions.  However,  if the rates are the same as for the sleptons, the effect
is about a 15\% decrease in $(\ohsq)^0/R$, leading to a similar bound
on $\mchi$ as in the CMSSM.

The potential significance of coannihilation effects had been emphasized
previously~\cite{gs,co2,dn}, particularly in the Higgsino LSP region. We
have shown that this can also be an important effect in the ${\tilde B}$
region, where the NLSPs include the ${\tilde \tau}_R$ and other
right-handed sleptons. In particular, we find in the context of the CMSSM
that the upper bound on the LSP mass quoted previously~\cite{upper} should
be increased by a factor of about two, to $\mchi\la600\gev$.

This observation has many potential ramifications, in particular for
searches for neutralino dark matter. Generically, in regions of
parameter space where coannihilation is important, a relic density
of astrophysical interest is now obtained for a smaller annihilation
cross section. This is likely to reduce the typical rates for
signatures of annihilations in the galactic halo. The corresponding 
elastic
scattering cross section is also likely to be reduced generically,
with a consequent suppression of signatures for scattering on nuclei and
capture followed by annihilation in the Sun or Earth.
These points are worthy of further study.

On the basis of the previously-quoted upper bound on the LSP mass, it
has been argued that the LHC is guaranteed to detect some supersymmetric
particles~\cite{CMS}, since it could reach out to $\m12 \sim 1200$ GeV
when
$\mb \sim m_{{\tilde \tau}_R}$. It is still true that the overwhelming
majority of the CMSSM parameter space allowed by cosmology can be
explored by the LHC, but there may be a narrow region 
of $m_0$ extending up to $\m12 \sim 1500$ GeV that is problematical.
This point is also worthy of further study.

\vskip .3in
\vbox{
\noindent{ {\bf Acknowledgments} } \\
\noindent  The work of K.O. was supported in part by DOE grant
DE--FG02--94ER--40823.  The work of T.F. was supported in part by DOE   
grant DE--FG02--95ER--40896 and in part by the University of Wisconsin  
Research Committee with funds granted by the Wisconsin Alumni Research  
Foundation.}


\newpage

\end{document}